\documentclass[twocolumn]{aastex631}

\shorttitle{Outer giants and missing planets}
\shortauthors{Lammers \& Winn}

\graphicspath{{./}{}}
\usepackage{amsmath}
\usepackage{comment}

\begin{document}

\title{The Gap-Giant Association: Are Planets Hiding in the Gaps?}


\author[0000-0001-9985-0643]{Caleb Lammers}
\affiliation{Department of Astrophysical Sciences, Princeton University, 4 Ivy Lane, Princeton, NJ 08544, USA}

\author[0000-0002-4265-047X]{Joshua N.\ Winn}
\affiliation{Department of Astrophysical Sciences, Princeton University, 4 Ivy Lane, Princeton, NJ 08544, USA}

\begin{abstract}

A handful of stars are known to host both an inner system of multiple transiting planets and an outer giant planet. These systems all feature a prominent gap between the orbits of two of the transiting planets, distinguishing them from typical multiplanet systems with more uniform orbital spacings. The reason for the association between inner gaps and outer giants is unknown. In this paper, we assess whether undiscovered planets might occupy these gaps in systems with outer giants. For each of the four relevant systems -- Kepler-48, Kepler-65, Kepler-90, and Kepler-139 -- we found that a typical small planet (${\sim}\,1$\,--\,$20\,M_\oplus$) could reside in the gap without inducing dynamical instability. However, in each case, the gravitational influence of the outer giant planet is insufficient to tilt the orbit of the hypothetical planet by enough to prevent transits, strongly disfavoring a proposed theory for the observed gap-giant association. The gaps might instead contain smaller, undetected planets (${\lesssim}\,1\,R_\oplus$), or be entirely devoid of planets.

\end{abstract}

\section{Introduction}
\label{sec:intro}

Although thousands of transiting planets are known, almost all of them orbit their stars within $1$~AU, and the outer regions of these systems remain largely unexplored. This ignorance stems primarily from the observational biases inherent to transit surveys (e.g., \citealt{Pepper2003, Gaudi2005, WinnPetigura2024}). Most of what is known about outer planetary companions comes from long-term radial velocity (RV) observations, which can detect outer giant planets after years of monitoring.\footnote{Throughout this paper, ``outer giant'' refers to planets with masses between $0.3$ and $13~M_\mathrm{Jup}$ and semi-major axes between $0.5$ and $10$~AU (see \citealt{Weiss2024}).} RV surveys show that outer giants occur around ${\sim}\,5$\,--\,$15$\% of nearby Sun-like stars \citep{Cumming2008, Rowan2016, Wittenmyer2016, Wittenmyer2020}. It remains unclear whether outer giants are more likely to coexist with inner planetary systems, or whether their presence influences the orbital architectures of inner systems \citep[see][]{Zhu&Wu2018, Bryan2019, Rosenthal2022, Bonomo2023, Bryan&Lee2024, VanZandt2025}.

\begin{figure}
\centering
\includegraphics[width=0.475\textwidth]{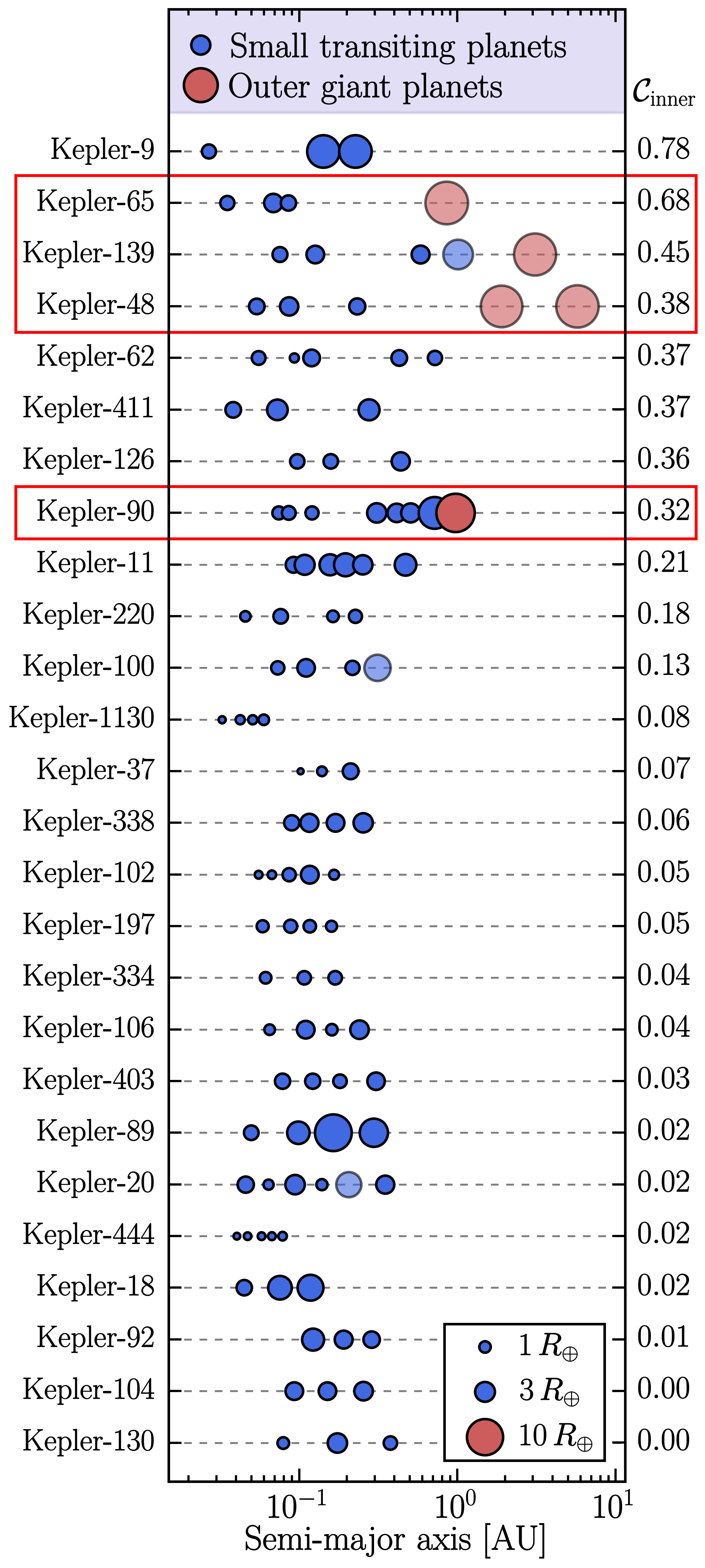}
\caption{Orbital architectures of systems in the Kepler Giant Planet Survey with three or more transiting planets. Inner planets are represented by blue circles, outer giant planets by red circles, and nontransiting planets by lighter-colored circles. Circle size is proportional to the square root of planet radius (as measured, or based on the \citealt{Chen&Kipping2017} mass-radius relationship if unmeasured). Systems are sorted by the gap complexity of the inner planetary system, $\mathcal{C_\mathrm{inner}}$, highlighting that the four systems known to have outer giants (red boxes) have unusually large orbital gaps (see also Figure~1 from \citealt{He&Weiss2023}).}
\label{fig:systems}
\end{figure}

Studies of the relationship between outer giants and inner planetary systems are often complicated by complex or poorly documented selection criteria for RV follow-up observations \citep{VanZandt2023}. To mitigate this problem, the Kepler Giant Planet Survey conducted nearly $3{,}000$ RV measurements targeting $63$ Sun-like stars with {\it Kepler}-detected transiting planets \citep{Weiss2024}. Among the $26$ systems in their sample with three or more transiting planets, four were found to host an outer giant (see Figure~\ref{fig:systems}). An intriguing pattern emerged: each of these four systems features a prominent orbital gap between two adjacent inner planets. Such gaps are unusual. Most of the {\it Kepler} multitransiting systems have nearly uniform logarithmic spacings \citep{Lissauer2011, Weiss2018}. This orbital spacing regularity, along with the similarity of planet sizes, causes most {\it Kepler} planetary systems to resemble ``peas in a pod'' in depictions such as Figure~\ref{fig:systems} \citep{Weiss2023}.

\citet{He&Weiss2023} quantified the tendency for systems with outer giants to have irregularly spaced inner planets using the gap complexity statistic, $\mathcal{C_\mathrm{inner}}$, a metric originally introduced by \citet{Gilbert&Fabrycky2020}. Low values of $\mathcal{C_\mathrm{inner}}$ are obtained when an inner planetary system has orbital distances forming a nearly log-uniform sequence, and large values are obtained in systems with orbital gaps. Hereafter, we will consider systems with $\mathcal{C_\mathrm{inner}}\,{>}\,0.3$ to feature a gap. \citet{He&Weiss2023} found that systems with outer giants feature larger values of $\mathcal{C_\mathrm{inner}}$ than those without outer giants at a modest level of statistical significance ($p\,{=}\,0.012$).

The cause of the ``gap-giant association'' is unclear. One hypothesis is that outer giants influence the formation or evolution of inner systems at early times. Most theoretical work on the effects of outer giants on planet formation has focused on whether outer giants enhance or inhibit the formation of inner planets, rather than on orbital spacings, and results have been inconclusive. Outer giants could hinder the formation of inner planets by preventing the inward flow of solids \citep[see, e.g.,][]{Izidoro2015, Lambrechts2019, Schlecker2021}, or they may enhance it by transporting planetesimals inwards via sweeping secular resonances \citep{Best2024}. A positive association could also arise if massive protoplanetary disks tend to form both outer giants and an abundance of solids closer to the star that ultimately form small planets \citep{Chachan&Lee2023, Bryan&Lee2024}. Alternatively, inner planet formation might proceed independently, aided by pressure bumps that promote the formation of planetesimals despite the reduction in pebble flux caused by the outer giant \citep{Bitsch&Izidoro2023}. These studies did not directly address orbital spacings. Recently, \citet{Kong2024} studied the role of giant planets in simulations of the giant impact phase and found that they tend to produce more evenly spaced orbits for the inner planets --- the opposite of the observed trend.

Another possible explanation for the gap-giant association is the dynamical influence of outer giants on the geometry of inner planetary systems. After formation, outer giants can induce mutual inclinations between inner planetary orbits through secular interactions, potentially tilting some orbits away from the line of sight and preventing them from transiting \citep[e.g.,][]{Becker&Adams2017, Lai&Pu2017, Mustill2017, Lammers&Winn2025}. This mechanism offers a simple and appealing explanation for the observed orbital gaps in systems with outer giants. This hypothesis was proposed by \citet{He&Weiss2023} and was recently advanced by \citet{Livesey&Becker2025}, who showed that the presence of a nearby massive companion tends to increase the gap complexity in simulated multiplanet systems.

To better understand the gap-giant association, we have scrutinized the orbital architectures of the four systems upon which the trend is based: Kepler-48, Kepler-65, Kepler-90, and Kepler-139 (see \citealt{He&Weiss2023} and Section~\ref{sec:other_systems} of this paper). We asked: (1) Could these systems host undetected planets in their gaps without becoming dynamically unstable? (2) Are the outer giants capable of tilting the orbits of the hypothetical gap planets by large enough angles to render them nontransiting? and (3) What observational limits can be placed on the existence of transiting planets in the gaps?

\section{Dynamical stability}
\label{sec:stability}

Previous studies have estimated that $60$ to $95$\% of {\it Kepler}'s multitransiting systems would become dynamically unstable if a several-Earth-mass planet were inserted between a pair of known planets \citep{Fang&Margot2013, Obertas2023}. This susceptibility to dynamical instability is largely due to the tight planet spacings of typical {\it Kepler} systems (see Figure~\ref{fig:systems}), which leaves little room for additional massive planets. Below, we explore whether the observed gaps in the Kepler-48, -65, -90, and -139 systems are sufficiently wide to accommodate a typical planet without inducing dynamical instability.

Before undertaking computationally intensive numerical simulations, we used approximate stability predictors to establish baseline expectations (see, e.g., \citealt{Chambers1996, Smith&Lissauer2009, Obertas2017, Tamayo2020, Lammers2024}). We assessed the stability of each system with a hypothetical planet placed in the observed gap. For simplicity, at this stage, we assumed initially circular and coplanar orbits, disregarded the influence of the outer giants, and placed the hypothetical planet at the midpoint of the gap. We varied the masses of the gap planets in this experiment. According to \texttt{SPOCK}, a machine-learning model trained to predict the long-term stability of compact multiplanet systems \citep{Tamayo2020, Cranmer2021}, all four systems could host an additional planet and remain stable for at least $10^9\,P_1$, where $P_1$ is the initial orbital period of the innermost planet. We found this to be true even if the hypothetical planet was several times more massive than the known inner planets. The empirical stability criteria of \citet{Obertas2017} and \citet{Lammers2024}, although calibrated for the special case of equal-mass and logarithmically spaced planets, can also be used for this purpose. We calculated the average planet mass and average period ratio for each system, with a hypothetical gap planet included, and provided these values as input to the empirical criteria. Both criteria predict that the four systems will remain stable over at least $10^9\,P_1$, even when the gap planets were massive.

This preliminary analysis suggested that these systems could stably support a hypothetical planet in the gap. However, we have neglected some potentially important effects: the gravitational influence of the outer giant, as well as any nonzero eccentricities and inclinations in the inner planetary system. These effects could make the systems less stable.

\subsection{Initial conditions}
\label{sec:ICs}

To assess stability more accurately, we created many realizations of each {\it Kepler} system based on the available observational constraints. Kepler-90 has a more delicate dynamical configuration than the other three systems that required more careful treatment. The initial conditions used in our dynamical simulations are described below.

\begin{table}
\centering
\caption{Properties of outer giants in Kepler-48 and Kepler-65 (from \citealt{Weiss2024}), Kepler-90 (from \citealt{Liang2021}), and Kepler-139 (from \citealt{Lammers&Winn2025}) used to initialize dynamical simulations (see Section~\ref{sec:stability}).}
\begin{tabular}{cccc}
 \hline
 Planet & Period [days] & Mass [$M_\oplus$] & Eccentricity\\
 \hline
 Kepler-48e & $998\,{\pm}\,4$ & $687\,{\pm}\,20$ & $0.00\,{\pm}\,0.03$\\
 Kepler-48f & $5220\,{\pm}\,374$ & $299\,{\pm}\,93$ & $0.02\,{\pm}\,0.14$\\
 Kepler-65e & $257.9\,{\pm}\,0.8$ & $217\,{\pm}\,22$ & $0.30\,{\pm}\,0.02$\\
 \tablenotemark{a}Kepler-90h & $331.6011$ & $203\,{\pm}\,5$ & $0.038\,{\pm}\,0.009$\\
 Kepler-139e & $1904\,{\pm}\,74$ & $378\,{\pm}\,44$ & $0.04\,{\pm}\,0.06$\\
 \hline
\end{tabular}
\raggedright
\tablenotetext{a}{Kepler-90h, the only giant planet in the sample known to transit, has a measured radius of $11.3\,{\pm}\,0.3\,R_\oplus$ \citep{Fulton&Petigura2018}.}
\label{table:measurements}
\end{table}

For Kepler-48 and Kepler-65, we sampled each planet's mass from independent normal distributions based on the means and standard deviations reported by \citet{Weiss2024}. For Kepler-139, we used updated constraints from \citet{Lammers&Winn2025}, which include a nontransiting planet of mass $35~M_\oplus$ on a $355$-day orbit. The initial orbital periods of the giant planets were drawn from normal distributions consistent with observational uncertainties (see Table~\ref{table:measurements}), while the periods of the inner planets were set equal to the best-fit values. Motivated by the typically low eccentricities of {\it Kepler} planets \citep[see, e.g.,][]{VanEylen&Albrecht2015, Hadden&Lithwick2017, Gilbert2025}, we drew the orbital eccentricities of the transiting planets from a Rayleigh distribution with scale parameter $\sigma_e\,{=}\,0.02$. Similarly, the orbital inclinations of the transiting planets were drawn from a Rayleigh distribution with scale parameter $\sigma_I\,{=}\,1.5^\circ$, consistent with observational constraints \citep{Fang&Margot2012, Fabrycky2014}. The mean longitudes, longitudes of pericenter, and longitudes of ascending node were drawn from uniform distributions ranging from $0$ to $2\pi$. The eccentricities of the outer giants are well constrained by the available RV data, and were sampled from the observational posteriors (Table~\ref{table:measurements}). However, their inclinations are unknown; following \citet{Masuda2020}, who inferred a typical misalignment of ${\sim}\,10^\circ$ between the orbits of outer giants and inner transiting planets, we sampled the initial inclinations of the outer giants from a Rayleigh distribution with $\sigma\,{=}\,10^\circ$. Orbital orientations were chosen randomly. Under these assumptions, we found that the Kepler-48, -65, and -139 systems remained stable for at least $10^9\,P_1$ (see Section~\ref{sec:nbody_sims} for details about our $N$-body simulations).

Kepler-90 differs from the other three systems in that the masses of its inner planets are poorly constrained. Additionally, when we initialized Kepler-90-like systems using the process described above, nearly every system went unstable before $10^9\,P_1$, even without a hypothetical gap planet. Specifically, in these tests, we adopted masses for planets g and h from the transit-timing analysis of \citet{Liang2021}, and drew the other planet masses from the \citet{Chen&Kipping2017} mass-radius relation. This finding indicates that the prescribed initial conditions are unrealistic for Kepler-90, consistent with the system's previously noted dynamical fragility \citep{Granados2018, Gaslac2024}. To ensure dynamical stability, we instead initialized all of Kepler-90's planets (including the transiting outer giant) on initially circular and coplanar orbits.

We then injected a hypothetical planet into each system's gap. To place the planet near the center of the gap, we drew its orbital period from a normal distribution centered on the geometric mean of the bordering planets' periods, $\sqrt{P_i\,P_{i\,{+}\,1}}$, with a standard deviation equal to $5$\% of this value. The injected planet's mass was sampled from a normal distribution with a mean and standard deviation set equal to the mean mass of the known inner planets in the system, with negative draws rejected and resampled. For Kepler-48, -65, and -139, we assigned eccentricities and inclinations from Rayleigh distributions with scale parameters $\sigma_e\,{=}\,0.02$ and $\sigma_I\,{=}\,1.5^\circ$, consistent with the other inner planets. For Kepler-90, the gap planet was placed on an initially circular and coplanar orbit. In both cases, the orbital angles were drawn randomly from uniform distributions. These choices yielded a variety of plausible gap planets that resembled the other inner planets, with $90$\% of their masses in the range $1$\,--\,$20\,M_\oplus$.

\subsection{$N$-body simulations}
\label{sec:nbody_sims}

To evaluate the long-term stability of our synthetic systems, we performed $N$-body integrations with the \texttt{WHFast} symplectic integrator \citep{Wisdom&Holman1991, Rein&Tamayo2015} from the open-source \texttt{REBOUND} package. We adopted the standard timestep of $dt\,{=}\,P_1/20$, and each simulation was run until either $10^9\,P_1$ had elapsed or the Hill spheres of two planets intersected (a reliable proxy for instability). For each system, we initialized and integrated $480$ realizations, with survival rates as follows:
\begin{itemize}
    \item Kepler-48: $97$\% (464/480) remained stable.
    
    \item Kepler-65: $87$\% (419/480) remained stable.

    \item Kepler-90: $96$\% (460/480) remained stable.
    
    \item Kepler-139: $62$\% (299/480) remained stable.
    
\end{itemize}
These high survival rates indicate that a gap planet could remain dynamically stable in each system for $10^9\,P_1$, which corresponds to $13$, $6$, $19$, and $20$~Myr for Kepler-48, -65, -90, and -139, respectively. The stable configurations appeared to be dynamically quiescent, with minimal eccentricity growth and nearly constant semi-major axes, suggesting stability on longer timescales. The relatively low survival fraction for Kepler-139 reflects, in part, the wide range of masses we considered for the gap planet ($m_3$). Perhaps counterintuitively, Kepler-139-like systems with low-mass gap planets were more prone to instability: those with $m_3\,{>}\,10\,M_\oplus$ had a $74$\% survival rate, compared to $54$\% for $m_3\,{<}\,10\,M_\oplus$.

To test whether the outer giants played a role in destabilizing the systems, we repeated the simulations with the outer giants omitted. The resulting survival fractions increased at least slightly in all cases ($100$\%, $91$\%, $99$\%, and $100$\% for Kepler-48, -65, -90, and -139, respectively), suggesting that the giant planets have a modest destabilizing effect. In the following section, we investigate whether the giants can trigger instabilities over longer timescales using secular perturbation theory.

\subsection{Secular integrations}
\label{sec:secular}

The dynamical influence of outer giant planets on inner planetary systems is often modeled using secular theory (see, e.g., \citealt{Laskar1990, Laskar1994, Lithwick&Wu2011, Boue&Fabrycky2014}). Secular theory is based on time-averaging the short-period terms in the Hamiltonian that governs the gravitational interaction between the planets. To second order in eccentricity and inclination, the resulting equations of motion decouple into two sets of linear first-order differential equations: the Laplace-Lagrange equations \citep{Murray&Dermott1999, Tremaine2023}. Retaining higher-order terms in eccentricity and inclination introduces the possibility of secular resonances, which can overlap and cause chaotic evolution \citep[e.g.,][]{Lithwick&Wu2011}. In the four systems studied here, the outer giants are distant and not near strong mean-motion resonances with the inner planets, making secular resonances the most plausible mechanism for inducing dynamical instability. Thus, we carried out secular integrations to explore stability on timescales beyond the reach of direct $N$-body simulations. These integrations are faster than $N$-body simulations because the shortest secular periods are typically much longer than the shortest orbital periods.

\begin{figure*}
\centering
\includegraphics[width=0.95\textwidth]{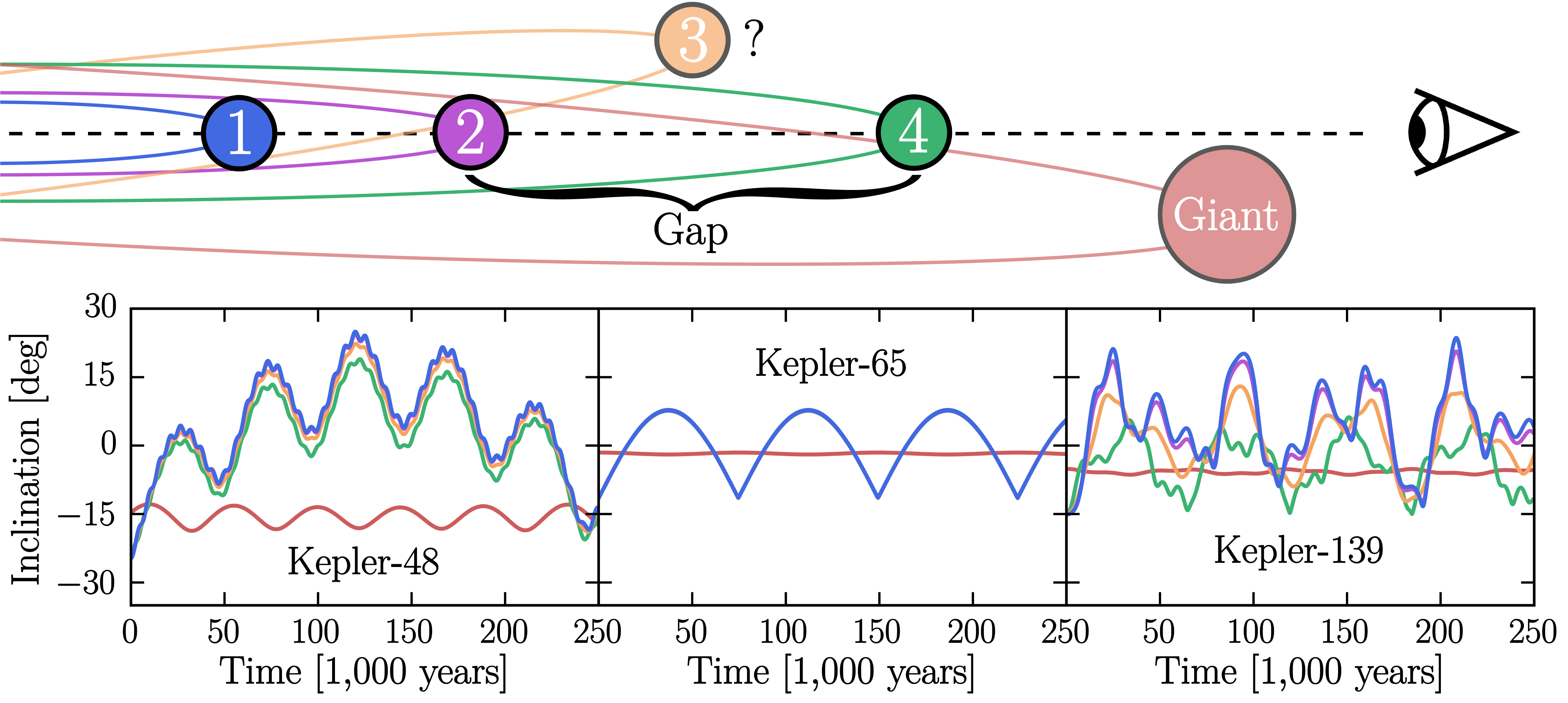}
\caption{Top: Schematic illustration of the ``secular inclination excitation'' hypothesis, in which the gravitational influence of an outer giant planet tilts the orbit of an inner planet away from the line of sight, creating an apparent gap in the sequence of transiting planets. Bottom: secular evolution in a representative realization of Kepler-48 (left), Kepler-65 (middle), and Kepler-139 (right). For clarity, the inclination curves of Kepler-48f and Kepler-139f are omitted. In Kepler-48 and Kepler-65, the outer giants are sufficiently distant from the inner systems to allow the inner planets to precess coherently, causing some of the planets' inclination curves to be hidden. In Kepler-139, the inner pair (planets 1 and 2) remains coupled, but the outer pair (planets 3 and 4) precesses more independently.}
\label{fig:schematic}
\end{figure*}

We conducted the secular simulations with the help of the \texttt{SecularSystemSimulation} class of the open-source \texttt{celmech} code \citep{celmech2022}, which algorithmically derives the equations of motion to arbitrary order in eccentricity and inclination. For this study, we included terms up to fourth order in both eccentricity and inclination. We integrated the resulting equations of motion using a Runge-Kutta method. Each system was integrated for $10^5\,T_{\mathrm{sec},\,1}$ with a timestep of $T_{\mathrm{sec},\,1}/20$, where $T_{\mathrm{sec},\,1}$ denotes the shortest secular period among the system's eccentricity and inclination eigenmodes.

A limitation of secular simulations is that they do not resolve close encounters between planets. To account for instabilities, we flagged a system as unstable whenever two planets' orbits satisfied the condition
\begin{equation}
    a_{i+1}(1\,{-}\,e_{i+1})\,{-}\,a_i(1\,{+}\,e_i)\,{<}\,r_{H,\,i+1}\,{+}\,r_{H,\,i}
\end{equation}
where $r_{H,\,i}\,{=}\,{a_i}\,(m_i/3M_\ast)^{1/3}$ is the Hill radius of the $i$th planet. This criterion, similar to that used by \citet{Lammers2024}, does not account for mutual inclinations. As a result, some systems may be flagged as unstable even though no close encounter has taken place, potentially resulting in underestimated instability timescales. The survival rates reported below should thus be interpreted as lower limits.

For Kepler-90, we initialized the system with all inner planets on circular, coplanar orbits, while assigning the outer giant an eccentricity and inclination sampled from the observational constraints reported by \citet{Liang2021}. This supplied the system with ${\sim}\,300$ times the critical angular momentum deficit (AMD) required for orbit crossings between planets $1$ and $2$, assuming the system's entire AMD were concentrated in that pair. In other words, our Kepler-90 configurations were ``AMD-unstable'' in the sense defined by \citet{Laskar&Petit2017}. The other three systems were initialized as described in Section~\ref{sec:ICs}, and were also AMD-unstable. We performed secular integrations for $480$ realizations of each system and recorded the following survival rates:
\begin{itemize}
    \item Kepler-48: $93$\% (448/480) remained stable.
    
    \item Kepler-65: $90$\% (433/480) remained stable.

    \item Kepler-90: $87$\% (419/480) remained stable.
    
    \item Kepler-139: $39$\% (185/480) remained stable.

\end{itemize}
Each system was found to be capable of hosting a gap planet without becoming secularly unstable over a timespan of at least $10^5\,T_{\mathrm{sec},\,1}$, corresponding to median durations of $43$, $16$, $18$, and $200$~Myr, for Kepler-48, -65, -90, and -139, respectively. Notably, for Kepler-90, $10^5\,T_{\mathrm{sec},\,1}$ was typically shorter than $10^9\,P_1$, so the secular integrations did not provide a computational advantage over direct $N$-body integrations.

Kepler-139 was significantly less stable than the other systems in our secular integrations. This difference cannot be explained solely by the system's relatively long secular period $T_{\mathrm{sec},\,1}$. Even over the shorter timespan of $10^4\,T_{\mathrm{sec},\,1}$ (${\sim}\,20$\,Myr), Kepler-139's survival fraction was only $53$\%, well below that of the other systems. We found the survival rate to be less sensitive to the gap planet's mass in the secular integrations than in the $N$-body simulations. Instead, the survival rate was most sensitive to the mass of the outer giant. Among simulations with $m_G\,{<}\,380\,M_\oplus$, the survival rate was $44$\%, while those with $m_G\,{>}\,380\,M_\oplus$ had a $33$\% survival rate. Overall, the survival rates are high enough to suggest that a planet could reside in Kepler-139's gap without destabilizing the system.

\section{Mutual transit analysis}
\label{sec:inc_sims}

Having established that a planet of comparable mass to the known inner planets could stably exist within the orbital gaps of the four relevant {\it Kepler} systems, we now consider whether perturbations from the outer giant could cause such a gap planet to become nontransiting (see illustration in Figure~\ref{fig:schematic}).

Before carrying out dynamical simulations, we established baseline expectations by referring to previous theoretical work on the influence of outer giants on the orbital orientations of inner companions \citep[e.g.,][]{Boue&Fabrycky2014, Becker&Adams2017, Huang2017, Lai&Pu2017}. A key diagnostic identified by these studies is the dynamical coupling parameter, defined as
\begin{align} 
\label{eps_def}
    \epsilon_{12} \equiv \frac{\Omega_{2G} - \Omega_{1G}}{\omega_{12} + \omega_{21}}~,
\end{align}
where $\Omega_{1G}$ is the precession rate of planet 1's angular momentum vector around that of the outer giant, and $\omega_{12}$ is the precession rate of planet 1's angular momentum vector around that of planet 2. These rates are calculated neglecting the effects of the other planets. The frequencies $\Omega_{2G}$ and $\omega_{21}$ are defined analogously for planet 2. The parameter $\epsilon_{12}$ quantifies how strongly the inclinations of planets $1$ and $2$ are coupled in response to perturbations from the outer giant. When $\epsilon_{12}\,{\gtrsim}\,1$, the planets precess independently at different rates, allowing their mutual inclination to grow to a value comparable to the giant's inclination (${\sim}\,\theta_G$). When $\epsilon_{12}\,{\lesssim}\,1$, the planets are strongly coupled and precess together, maintaining a smaller mutual inclination on the order of $\epsilon_{12}\,\theta_G$. Following \citet{Lai&Pu2017}, we estimated $\epsilon_{12}$ as
\begin{align} 
\label{eps_coupling}
\begin{split} 
    \epsilon_{12}\,{\approx}\,&\left(\frac{m_G}{m_2}\right) \left(\frac{a_2}{a_G}\right)^{\!\!3} \left[\frac{3a_1/a_2}{b_{3/2}^{(1)}(a_1/a_2)}\right]\\ &\left[\frac{(a_2/a_1)^{3/2} - 1}{(m_1/m_2)(a_1/a_2)^{1/2} + 1}\right]
\end{split}
\end{align}
where $b_{3/2}^{(1)}(\cdot)$ is the standard Laplace coefficient.

\begin{figure*}
\centering
\includegraphics[width=0.95\textwidth]{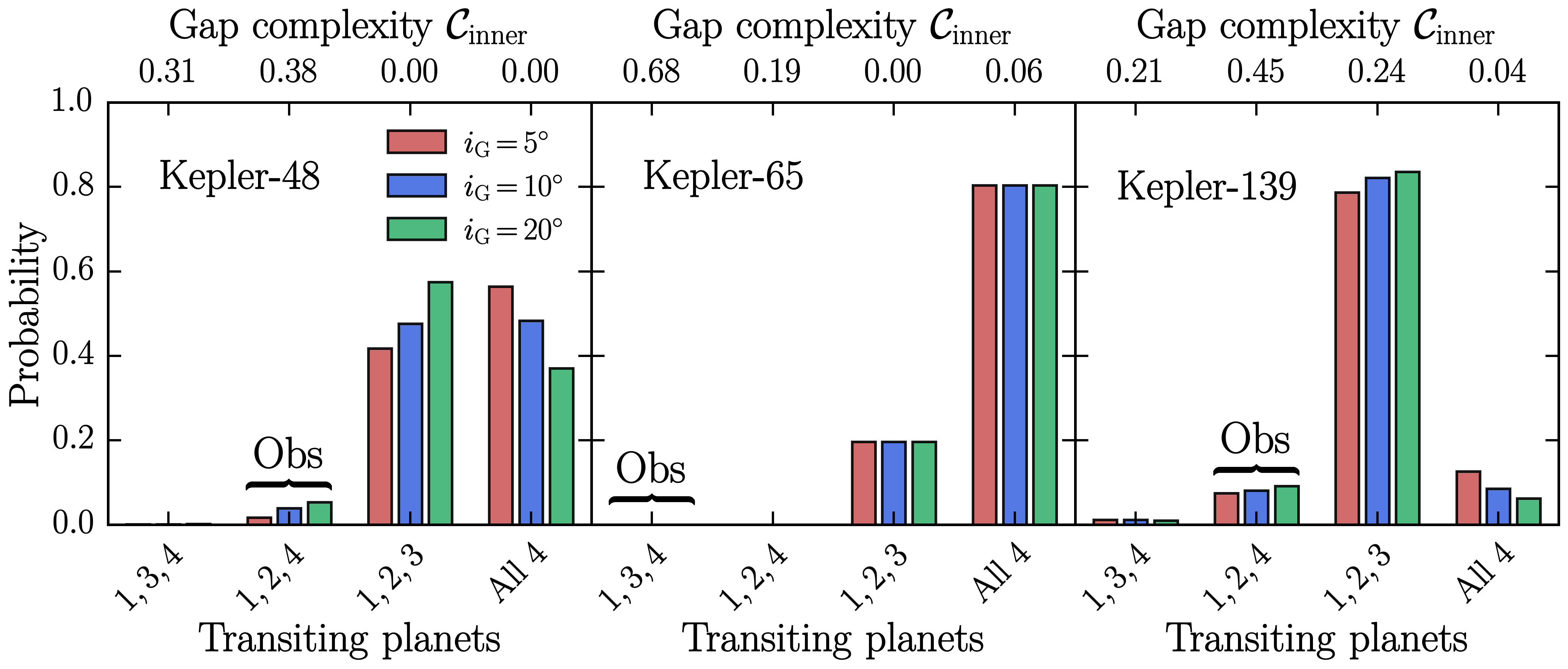}
\caption{Probability of various combinations of planets transiting simultaneously, assuming at least three planets transit. Each system includes a hypothetical planet in the gap, resulting in four inner planets (numbered as in Figure~\ref{fig:schematic}). The configurations corresponding to the actual {\it Kepler} observations are highlighted. Transit probabilities were computed by averaging over viewing angles and time-averaging over $10^6$\,years of secular evolution (see Section~\ref{sec:inc_sims}). Configurations in which planets 2, 3, and 4 transit without planet 1 are vanishingly rare and are excluded from the bar plots. In all three systems, the observed configuration was reproduced infrequently by the simulations ($5.3$\%, $0.001$\%, and $9.1$\% for Kepler-48, -65, and -139, respectively), indicating that secular perturbations from the outer giant are unlikely to explain the orbital gaps.}
\label{fig:transit_probabilities}
\end{figure*}

We initialized many realizations of each system using the process described in Section~\ref{sec:ICs}, and calculated the coupling parameter $\epsilon$ between each hypothetical gap planet and its nearest neighbors. The median values of $\epsilon$ were found to be approximately $0.02,~0.001,~0.2,~\mathrm{and~}0.07$ for Kepler-48, -65, -90, and -139, respectively. For Kepler-48, we disregarded the influence of planet f (the second outer giant) because it is both more distant and less massive than planet e, and therefore dynamically less significant.

For Kepler-65, the extremely low values of $\epsilon\,{\approx}\,0.001$ indicate strong dynamical coupling between a gap planet and its neighbors. This protects the inner system from developing mutual inclinations due to the outer giant, preventing a gap between transiting planets. Even if the outer giant were on a highly inclined orbit, the induced mutual inclinations among the inner planets would remain below ${\sim}\,1^\circ$ ($0.003$~rad). A similar conclusion applies to Kepler-48, where the values of $\epsilon\,{\approx}\,0.02$ imply that only a drastically inclined outer giant (${\gtrsim}\,50^\circ$) could induce mutual inclinations of ${\sim}\,1^\circ$. The intermediate coupling strengths found for Kepler-139 ($\epsilon\,{\approx}\,0.07$) suggest a more nuanced evolution that warranted further investigation, as described below (Section~\ref{sec:transit_probs}).

In the Kepler-90 system, the relatively high coupling parameters ($\epsilon\,{\approx}\,0.2$) imply that the outer giant could induce significant mutual inclinations if it were on a modestly inclined orbit. However, the outer giant is known to transit, strongly suggesting that its orbit is well-aligned with those of the inner system. While a large misalignment between the outer giant and the inner system is possible if the outer giant's longitude of ascending node differs from those of the other planets, such a configuration is statistically unlikely. If the giant planet's orbit were randomly oriented relative to the line of sight, its geometric transit probability would be only $0.55$\%. On this basis, we reject the secular inclination excitation hypothesis for the observed gap in Kepler-90 and exclude this system from the discussion below.

\subsection{Transit probabilities}
\label{sec:transit_probs}

For a quantitative assessment of the secular inclination excitation hypothesis, we performed Monte Carlo transit probability calculations for the Kepler-48, -65, and -139 systems. Each system was initialized with a hypothetical gap planet, and transit probabilities were tracked throughout the system's secular evolution.

Transit probabilities were calculated using a Monte Carlo method similar to that of \cite{Ragozzine&Holman2010}, \cite{Lissauer2011b}, and \cite{Lammers&Winn2025}. We generated many realizations of each system, assigned random viewing directions, and recorded which planets would be observed to transit. Specifically, we assigned the inclination $i_\mathrm{ref}$ of the system's reference plane by drawing $\cos\,i_\mathrm{ref}$ from a uniform distribution on the interval [$-1$, $1$]. For each realization, a planet was deemed transiting whenever the projected sky-plane distance $|a_p \cos i_{p,\mathrm{sky}}|$ was smaller than $R_\ast$, where 
\begin{equation} 
\label{mutual_inc}
\cos i_{p,\mathrm{sky}}\,{=}\,\cos i_\mathrm{ref} \cos i_p \,{+}\,\sin i_\mathrm{ref} \sin i_p \cos \Omega_p.
\end{equation}
The longitudes of the ascending node, $\Omega_p$, were drawn from a uniform distribution between $0$ and $2\pi$.

We initialized the systems following the procedure described in Section~\ref{sec:stability}, except that the gap planet was always placed at the midpoint of the gap, for simplicity. The inner planets were initialized on coplanar orbits, while the outer giant was assigned an inclination $i_\mathrm{G}$ relative to the inner system. By assuming initial coplanarity of the inner system, we isolated the dynamical effect of the outer giant; beginning with mutual inclinations among the inner planets introduces the possibility of observing an apparent gap unrelated to the giant planet's influence, complicating the analysis. We have confirmed that introducing small initial mutual inclinations (${\sim}\,1^\circ$) does not meaningfully alter the transit probabilities. The gap planet's mass in each system was sampled from a normal distribution whose mean and standard deviation were both set equal to the mean mass of the inner planets.

For Kepler-48, -65, and -139, we created $300{,}000$ realizations of the system, integrated the secular (Laplace-Lagrange) equations for each system for $10^6$~years (as shown in the bottom panel of Figure~\ref{fig:schematic}), and recorded which planets would be seen to transit. The Laplace-Lagrange equations of motion were derived with the help of the \texttt{LaplaceLagrangeSystem} class from the \texttt{celmech} package \citep{celmech2022}. We repeated this experiment for three different inclinations of the outer giant: $5^\circ$, $10^\circ$, and $20^\circ$. To match observations, we focused only on outcomes in which at least three planets transit. Figure~\ref{fig:transit_probabilities} shows the probabilities for different planet combinations that transit simultaneously, along with the corresponding gap complexities $\mathcal{C}_\mathrm{inner}$.\footnote{For Kepler-139, we included the nontransiting planet Kepler-139f in the simulations, but did not restrict the systems to those in which planet f avoids transiting, to simplify the interpretation of the results.} The results for each system are summarized below:
\begin{itemize}
    \item Kepler-48: When the outer giant's inclination is low ($i_\mathrm{G}\,{\lesssim}\,5^\circ$), the most common case is that all four inner planets are transiting (with a probability of ${\sim}\,60$\%). As $i_\mathrm{G}$ increases, it becomes increasingly likely that the outermost planet, d, avoids transiting. Observing a configuration with a gap (i.e., $\mathcal{C_\mathrm{inner}}\,{>}\,0.3$) requires that planets 1 and 4 transit but either 2 or 3 does not transit. This is rare, regardless of the inclination of the outer giant. Even if $i_\mathrm{G}\,{=}\,20^\circ$, the chance of seeing a gap is only $5.5$\%. Specifically, there is a $5.3$\% chance that planets 1, 2, and 4 transit and a $0.2$\% chance that planets 1, 3, and 4 transit.
    
    \item Kepler-65: The outer giants in this system have little impact on the transit probabilities of the inner system, due primarily to their wide orbits. The inner system behaves as a tightly coupled group with inclinations that evolve together (see the middle panel of Figure~\ref{fig:schematic}). The most likely configuration (with ${\sim}\,80$\% probability) is that all four inner planets are seen transiting. The ${\sim}\,20$\% chance that planet 4 is the only nontransiting planet is attributed to the lower geometric transit probability of the outermost planet, not dynamical perturbations from the outer giant. The probability of observing a gap, as in the actual configuration, is only $0.001$\%.

    \item Kepler-139: This system stands out
    because of the especially large gap between planets 2 and 4, and its relatively close-orbiting outer giant (see Figure~\ref{fig:systems}). The weaker coupling between planet 4 and the other inner planets leads to inclination oscillations that are more independent (see the green curve in Figure~\ref{fig:schematic}), making it relatively likely that planet 4 fails to transit when planets 1, 2, and 3 are all transiting. As a result, a random observer is much more likely to see transits of only planets 1, 2, and 3 (Figure~\ref{fig:schematic}). It is possible, although unlikely, that a random observer would see a gap in the system. For $i_\mathrm{G}\,{=}\,20^\circ$, we found a $9.1$\% chance of observing a gap.

\end{itemize}

In summary, the secular inclination excitation hypothesis cannot adequately account for the observed gaps in each of the four relevant {\it Kepler} systems.

\section{Limits on undetected planets}
\label{sec:obs_limits}

If transiting planets do exist within the gaps of the four {\it Kepler} systems we have studied, how small must they be to have evaded detection? To address this question, we turned to the {\it Kepler} team's study of the efficiency of their planet detection pipeline \citep[e.g.,][]{Christiansen2016, Thompson2018, Christiansen2020}. We used the \texttt{KeplerPORTs} code \citep{Burke&Catanzarite2017} to estimate detection efficiencies for the four systems.\footnote{available at \url{https://github.com/nasa/KeplerPORTs}} Provided the relevant stellar properties, \texttt{KeplerPORTs} calculates detection efficiencies as a function of planet radius and orbital period based on the ``multi-event statistic.'' This quantity is proportional to the transit signal-to-noise ratio and takes into account the details of the transit search pipeline. The model incorporates the dependence of the photometric noise on the averaging timescale (the ``combined differential photometric precision slope''; \citealt{Christiansen2012}). The multi-event statistic is calibrated by comparison with inject-and-recover experiments (see \citealt{Burke&Catanzarite2017} for more details).

\begin{figure}
\centering
\includegraphics[width=0.45\textwidth]{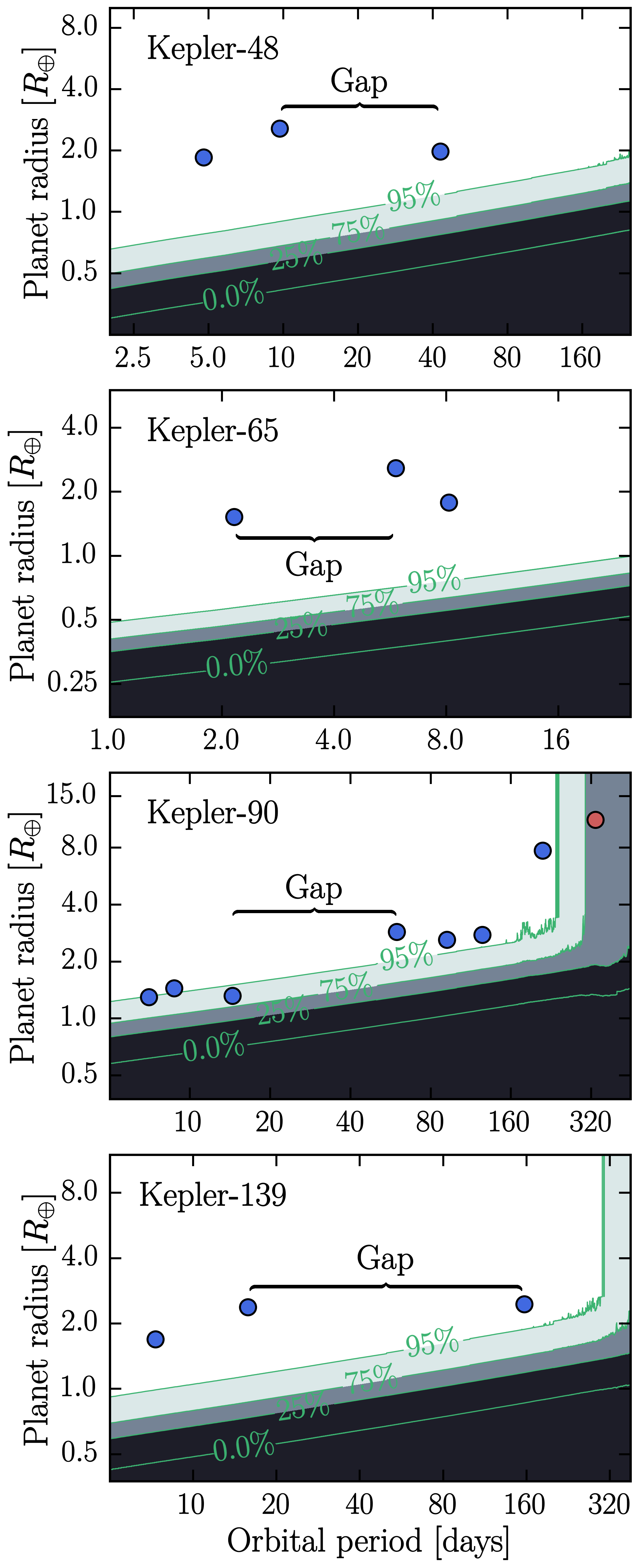}
\caption{Transit detection efficiency contours for Kepler-48, -65, -90, and -139, estimated with the help of the \texttt{KeplerPORTS} software. To have avoided detection in Kepler-48, -65, and -139, a hypothetical gap planet must be somewhat smaller than the known planets (${\lesssim}\,0.7\,R_\oplus$, ${\lesssim}\,0.5\,R_\oplus$, and ${\lesssim}\,1.0\,R_\oplus$, respectively). For Kepler-90, a gap planet that is of comparable size to the known planets could have evaded detection.}
\label{fig:DE_contours}
\end{figure}

The resulting detection efficiency contours are shown in Figure~\ref{fig:DE_contours}. For Kepler-48, -65, and -139, all known transiting planets lie above the $95$\% confidence detection threshold.\footnote{Detection efficiencies may be overestimated for large stars like Kepler-65 because the maximum transit duration considered by the pipeline ($15$~hours) can be exceeded. This does not affect our analysis of Kepler-65, however, because it was restricted to short orbital periods.} For a planet in the gap of one of these systems to go undetected, it must therefore be somewhat smaller than the known planets.

By contrast, in Kepler-90, all of the known planets hover near the detection threshold. Indeed, the eighth planet, Kepler-90i ($P_\mathrm{orb}\,{=}\,14$\,days, $R_p\,{=}\,1.3\,R_\oplus$), was missed by three teams that analyzed the system \citep{Cabrera2014, Rowe2014, Schmitt2014} and was only identified later with the help of machine-learning tools \citep{Shallue&Vanderburg2018}. At the midpoint of Kepler-90's gap, a planet as large as Kepler-90b or i ($1.3\,R_\oplus$) would be detected less than $50$\% of the time. Thus, the gap in the Kepler-90 system could plausibly contain a transiting planet that is slightly too small to have been detected.

As noted in Section~\ref{sec:intro}, typical {\it Kepler} multiplanet systems feature relatively uniform planet sizes (see Figure~\ref{fig:systems} and \citealt{Lissauer2011, Weiss2018}). For a gap planet to go undetected, how much must Kepler-48, -65, and -139 systems deviate from this trend? To answer this question, we identified the radius required for a detection probability below $50$\% at the midpoint of each gap. The results were ${\lesssim}\,0.7\,R_\oplus$, ${\lesssim}\,0.5\,R_\oplus$, and ${\lesssim}\,1.0\,R_\oplus$ for Kepler-48, -65, and -139, respectively. To contextualize these values, we computed the radius dispersion, $\sigma_{R_p}/\bar{R_p}$, that would result from inserting a planet this small into each system. For Kepler-48, -65, and -139, the resulting dispersion would be ${\gtrsim}\,0.45$, ${\gtrsim}\,0.55$, and ${\gtrsim}\,0.36$, respectively. We compared this with all {\it Kepler} systems that contain three or more confirmed planets.\footnote{Using data from the NASA Exoplanet Archive, \url{https://exoplanetarchive.ipac.caltech.edu}, accessed on April 2nd, 2025.} The median radius dispersion among these systems is $0.26$, with a standard deviation of $0.20$. Based on these results, including a marginally detectable gap planet in the Kepler-48, -65, and -139 systems would place them in the top $17$\%, $10$\%, and $30$\% of all {\it Kepler} systems sorted by radius dispersion. Thus, the radius dispersions of these systems would make them only slightly atypical. Moreover, because such small planets would often have gone undetected in many other {\it Kepler} systems, the population-level estimate of the radius dispersion is likely to be underestimated relative to the true value (see also \citealt{Murchikova&Tremaine2020, Zhu2020}). Based on the handful of detections, it is expected that planets smaller than $1~R_\oplus$ are common in systems that host super-Earths or sub-Neptunes \citep{Hsu2019, Qian&Wu2021}. The gap-giant association may stem from an association between outer giants and inner sub-Earths.

Other authors have noticed correlations between planet sizes and the presence of gaps. \citet{Rice2024} reported that {\it Kepler} systems with $\mathcal{C}_\mathrm{inner}\,{>}\,0.35$ feature fewer sub-Earths and sub-Neptunes, but more super-Earths, than systems with $\mathcal{C}_\mathrm{inner}\,{<}\,0.165$ (see also \citealt{Chance2024, Goyal&Wang2024}). The interpretation of this finding is somewhat unclear, but it seems possible that many systems with gaps host undiscovered sub-Earths.

\section{Discussion}
\label{sec:discussion}

\subsection{Other multiplanet systems with outer giants}
\label{sec:other_systems}

In addition to the four systems from \citet{He&Weiss2023}, we found two other systems in the NASA Exoplanet Archive that are known to host three small transiting planets ($R\,{<}\,4\,R_\oplus$) and an outer giant planet: Kepler-167 and HD~191939. The outer giant in Kepler-167 has an orbital period of $1071$~days and is known to transit \citep{Kipping2016}. The inner planets in this system are located far from the giant, with orbital periods of $4.4$, $7.4$, and $22$~days, and feature a gap between planets c and d ($\mathcal{C}_\mathrm{inner}\,{=}\,0.36$).

HD~191939 is a bright ($V\,{=}\,9.0$) G-type star which was observed as part of the Transiting Exoplanet Survey Satellite's primary mission. These observations revealed three transiting sub-Neptune-sized planets \citep{BadenasAgusti2020}, and subsequent RV follow-up led to the discovery of a giant planet ($M\,{=}\,0.34\,M_\mathrm{Jup}$) on a $101$-day orbit \citep{Lubin2022}, as well as two longer-period planets \citep{OrellMiquel2023, Lubin2024}. HD~191939's three inner planets have orbital periods of $8.9$, $29$, and $38$~days, resulting in a large gap between planets b and c ($\mathcal{C}_\mathrm{inner}\,{=}\,0.85$). The fact that both Kepler-167 and HD~191939 feature gaps provides additional evidence for the gap-giant association, although we recommend caution when trying to perform statistical inferences based on systems that were individually selected for RV follow-up (like HD~191939).

Could secular dynamics be responsible for the gaps in these two systems? Kepler-167's giant is known to transit, making its orbit unlikely to be misaligned with the inner planets. Furthermore, even if it were grossly misaligned, the giant is far enough from the inner planets to render the secular inclination excitation hypothesis untenable. HD~191939 appears more promising, at first glance, due to the relative proximity of the giant to the inner system. To test the secular dynamics hypothesis, we carried out the transit probability experiments described in Section~\ref{sec:transit_probs} for this system. We found, once again, that the outer giant's dynamical influence is unlikely to produce a gap. In this case, a randomly oriented observer has just an $8$\% chance of observing the system when a gap is present.

Beyond systems of multiple transiting planets, systems with inner planets detected via the Doppler method may be useful for understanding the gap-giant association. Because planets can be detected with RV observations even when they are not transiting, a natural prediction of the undetected-planet scenario is that RV-detected inner systems accompanied by outer giants should not feature gaps. There are two systems listed in the NASA Exoplanet Archive that host three RV-detected low-mass inner planets ($m\,{<}\,30\,M_\oplus$) and an outer giant: HD~164922 and HD~219134. HD~164922 is a bright ($V\,{=}\,7.0$) G-type star that has long been known to host an outer giant \citep{Butler2006} and was later found to host three low-mass inner planets \citep{Fulton2016, Benatti2020, Rosenthal2021}. The system features a gap ($\mathcal{C}_\mathrm{inner}\,{=}\,0.35$) between the first and second planets, which are further apart than the second and third planets ($P_\mathrm{orb}\,{=}\,12$,~$42$,~and~$76$\,days). The existence of a gap supports the gap-giant association and hints that some gaps may be devoid of planets, although the system may still host undetected low-mass planets.

HD~219134 is a bright ($V\,{=}\,5.6$) K-type star that was reported to host three inner planets and an outer giant by \citet{Motalebi2015}. Soon after, \citet{Vogt2015} announced the discovery of two additional low-mass planets in the system. HD~219134 b and c are known to transit \citep{Motalebi2015, Gillon2017}, and the detection of the giant (planet h) is relatively secure, but the existence of planets f, d, and g has been challenged because their orbital periods are near harmonics ($1/2$, $1$, and $2$, respectively) of the stellar rotation period \citep{Johnson2016, Folsom2018}. Assuming all planets are real, the HD~219134 system is relatively evenly spaced ($\mathcal{C}_\mathrm{inner}\,{=}\,0.07$) with no gap, defying the gap-giant association. A larger sample of RV-detected inner planetary systems would help improve our understanding of the gap-giant trend.

\subsection{Systems with gaps and no outer giants}
\label{sec:gap_no_OGs}

There are a handful of systems in Figure~\ref{fig:systems} with prominent gaps and no known outer giant (Kepler-9, -62, -411, and -126). For Kepler-9, -62, and -126, the available RV data rule out the existence of giants with similar mass and separation to those seen around Kepler-48, -65, -90, and -139 (see \citealt{Weiss2024} for quantitative upper limits). Thus, the presence of a giant does not seem to be a \emph{necessary} condition for the presence of a gap in the inner system. There are many possible causes for these gaps, including undetected planets (transiting planets with sizes below {\it Kepler}'s sensitivity limits or planets on inclined orbits) and planet-planet scattering \citep[e.g.,][]{Hansen&Murray2013}.

\subsection{Conclusion}
\label{sec:conclusion}

The influence of outer giant planets on inner planetary systems remains poorly understood. The handful of systems with a known outer giant and multiple transiting planets has revealed an intriguing trend: these systems are more likely to feature a large gap between the orbits of two of the transiting planets (see Figure~\ref{fig:systems}). The underlying cause of this correlation is unclear. Before our work, it seemed plausible that the outer giants' gravitational influence induced mutual inclinations among the orbits of the inner planets, causing some to become nontransiting and thereby creating apparent gaps (see \citealt{He&Weiss2023, Livesey&Becker2025}).

In this paper, we critically evaluated the possibility that the gaps observed in the four relevant systems could be due to undetected planets. Through dynamical simulations, we found that each of the four systems could stably host an additional planet in its gap. However, in each system, the outer giant's dynamical influence fails to excite sufficient mutual inclinations to prevent the gap planet from transiting. Although it is possible, in principle, for secular dynamics to produce gaps in generic multiplanet systems \citep{Livesey&Becker2025}, we concluded that this mechanism is unlikely to be responsible for the gaps in the observed systems.

We also explored whether the gaps could be explained by undetected transiting planets below {\it Kepler}'s sensitivity threshold. We found this hypothesis to be plausible, in that hosting such a planet would not cause the systems to deviate dramatically from the radius uniformity in typical {\it Kepler} systems. This raises the possibility that the observed association between gaps and outer giants is coincidental. A larger sample of multitransiting planets with outer giants would help to assess its significance. If the association proves to be real, our analysis indicates that dynamical evolution is unlikely to be responsible for the gaps. Instead, the connection may point to processes that took place during planet formation. Perhaps, the influence of outer giants on the inward flow of material promotes the formation of sub-Earths or imprint gaps between the inner planets. Alternatively, the dissipation of the protoplanetary disk may establish secular resonances between inner planets and outer giants, exciting mutual inclinations.

In the coming years, ongoing RV surveys and precise time-series astrometry from {\it Gaia} will expand the sample of systems with outer giants and multiple inner planets. These data will help to clarify whether the gap-giant association holds (and, if so, why), and may reveal new links between the architectures of inner and outer planetary systems.

\section{Acknowledgments} 
\label{sec:acknowledgments}

We thank the anonymous referee for a timely and thorough report, along with Juliette Becker, Daniel Fabrycky, Jeremy Goodman, Joseph Livesey, Alexander Thomas, Lauren Weiss, and Wenrui Xu for helpful discussions. We are pleased to acknowledge that the work reported in this paper was substantially performed using the Princeton Research Computing resources at Princeton University, which is a consortium of groups led by the Princeton Institute for Computational Science and Engineering (PICSciE) and the Office of Information Technology’s Research Computing.


\bibliography{refs}{}
\bibliographystyle{aasjournal}

\end{document}